\documentclass[aps,preprint,floats,showpacs,superscriptaddress,tightenlines]{revtex4}
\usepackage{epsfig}
\begin{document}
\title{Residual Chiral Symmetry Breaking in Domain-Wall Fermions}
\author{Chulwoo Jung}

\affiliation{
Department of Physics,
University of Maryland,
College Park, MD 20742, USA 
{~}}

\author{Robert G. Edwards}
\affiliation{
Jefferson Lab,
12000 Jefferson Avenue,
MS 12H2,
Newport News, VA 23606, USA 
{~}}

\author{Xiangdong Ji}
\author{Valeriya Gadiyak}
\affiliation{
Department of Physics,
University of Maryland,
College Park, MD 20742, USA 
{~}}

\preprint{UMD PP\#01-002}
\preprint{JLAB-THY-00-27}
\date{DOE/ER/40762-210~ July 2000}
\bigskip

\begin{abstract}
We study the effective quark mass induced by the 
finite separation of the domain walls in the domain-wall 
formulation of chiral fermion as the function of the size of
the fifth dimension ($L_s$), the gauge coupling $\beta$ and  
the physical volume $V$. We measure the
mass by calculating the small eigenvalues
of the hermitian domain-wall Dirac operator ($H_{\rm DWF}(m_0))$ in 
the topologically-nontrivial quenched $SU(3)$ 
gauge configurations. We find that the induced
quark mass is nearly independent of the physical volume, 
decays exponentially as a function of $L_s$, and has 
a strong dependence on the size of quantum fluctuations
controlled by $\beta$. The effect of the choice of 
the lattice gluon action is also studied.
\end{abstract}
\maketitle

Simulating massless or near-massless fermions on a lattice 
is a serious challenge in numerical quantum field theory. 
The origin of the difficulty can be traced to the 
well-known no-go theorem first shown by Nielsen and
Ninomiya, which states that one can not write down a local, 
hermitian, and chirally-symmetric lattice fermion action 
without the fermion doubling problem \cite{nn}. 
Hence to have chirally-symmetric fermions on a lattice, 
one must use nonlocal actions in which the coupling between 
lattice sites do not identically vanish even when the separation
becomes large. This implies that the lattice simulation with
chiral fermions is necessarily more expensive than, e.g, 
the standard Wilson fermion in which only the nearest neighbor
coupling is involved.

One of the lattice chiral fermion formalisms that have been studied 
extensively in recent years is the domain-wall fermion, 
first formulated by D. Kaplan \cite{kaplan} and later modified for 
realistic lattice simulation by Shamir \cite{shamir}. In the 
domain-wall construction, one introduces an extra fifth dimension 
$s$ with a finite extension. After discretization, 
the fifth direction
has $L_s$ number of lattice sites. If we put the same
four-dimensional gauge configuration on every four-dimensional 
$s$ slices, the five-dimensional massive theory admits a 
four-dimensional effective theory in which a left-handed chiral 
fermion lives near the $s=0$ slice and a right-handed one near  
$s=L_s-1$. Integrating out the heavy modes 
\cite{neuberger,kikukawa,borici,edwards1}, one
obtains an effective four-dimensional chiral theory 
in the limit of $L_s\rightarrow \infty$. 

For finite $L_s$, however, the two chiral modes can couple 
to produce an effective quark mass. Strong gauge field 
fluctuations can induce rather strong coupling, and hence rather 
large quark mass. This quark mass is expected to decrease
exponentially as $L_s\rightarrow \infty$ with possible power law 
corrections.
To gain a quantitative understanding
of this induced mass, Columbia and other groups have considered 
several different ways to measure it. One way is to study
the behavior of pion mass as the function of the wall
separation $L_s$ \cite{chen,aoki}. The problem with this 
is that the pion mass may not vanish as $L_s\rightarrow \infty$
due to the finite volume effects, and hence the non-zero
pion mass cannot be entirely attributed to the quark mass effect. 
Another way is to use the Gell-Mann--Oakes--Renner relation 
\cite{fleming,vranas}. 
However, this relation assumes a more complicated form in
the domain-wall fermion formalism. The effective mass can be also 
measured by studying the anomalous contribution in 
the axial Ward--Takahashi(WT) identity \cite{aoki,cppacs,rbc}. 

In Ref. \cite{gjj}, we proposed to measure the induced effective
quark mass by considering the eigenvalue of the hermitian 
domain-wall Dirac operator in the presence of the instanton background. 
In the $L_s\rightarrow \infty$ limit, the lattice version of
the Atiyah-Singer theorem \cite{at,overlap} guarantees 
the existence of exact
zero modes. For finite $L_s$, the lowest eigenvalues of the
Dirac operator are not zero. We take the average of these
would-be-zero eigenvalues as the effective quark mass. 
Our previous study for 150 gauge configurations at $\beta=6.0$
yields an effective quark mass $0.0074\pm 0.0007$ at $L_s=8$, 
$0.0022\pm 0.0003$ at $L_s=12$, and $0.0008\pm 0.00013$ at
$L_s=16$. The result is qualitatively consistent with 
those obtained from other methods \cite{aoki,cppacs,rbc}.

In this paper, we report a more systematic study of the effective
quark masses along this direction. In particular, we would like
to understand the effects of different lattice sizes, 
the coupling constant $\beta$, and the form of gluon actions. 
Crucial to the size of the effective mass is the near-zero 
eigenvalues of the hermitian Wilson-Dirac operator at 
the fixed domain-wall height $m_0$. These near-zero 
eigenvalues require large $L_s$ to project out the correct 
effective Dirac operator. While we find that the effective mass is 
largely independent of the lattice size, it is a sensitive
function of $\beta$ and the form of gluon actions. 
For a strong coupling (small $\beta$) the effective low-energy 
theory is recovered only at very large $L_s$. 
We find that the exponential decay rate for the effective mass is well
described by the density of zero eigenvalues of the hermitian
Wilson-Dirac operator.
For an improved gluon action, the quantum fluctuations are
strongly reduced and hence the domain-wall formalism
works much more efficiently.  

The domain-wall fermion was first introduced by D. Kaplan
\cite{kaplan}, based on an interesting observation that 
under certain conditions a five-dimensional 
massive fermionic theory has an effective 
four-dimensional massless fermion. For practical lattice simulations,
Shamir considered a five-dimensional lattice having
a finite extension in the fifth direction $s$ ($s=0, 1, 
\cdots L_s-1$). The lattice domain-wall fermion 
action is a five-dimensional Wilson 
action at $r=1$,  
\begin{eqnarray}
S_{\rm DWF} &=& \sum \overline{\psi} D_{\rm DWF}(m_0)\psi = 
   -\sum_{x,s=0}^{L_s-1}
  \overline\psi_{x,s}\biggl[ (m_0-5)\psi_{x,s}
 +P_R\psi_{x,s-1} + P_L\psi_{x,s+1} \nonumber \\
&&  + \frac12\sum_{\mu}
((1-\gamma_\mu)U_{x,\mu}\psi_{x+\mu,s}+
(1+\gamma_\mu)U^{\dagger}_{x-\hat{\mu},\mu} \psi_{x-\hat{\mu},s})
\biggr] \ , 
\label{action}
\end{eqnarray}
where $m_0$ is the negative of the conventional Wilson mass and
$P_{R,L}=\frac12(1\pm\gamma_5)$ is the chiral projection operators. 
The physical significance of $m_0$ is
the ``height of the domain wall". Like the inverse of 
the lattice spacing $1/a$ (taken to be 1), it acts as a heavy mass 
scale of the theory. The gauge fields live on the four-dimensional 
sub-lattices and are independent of $s$. Dirichlet 
boundary conditions are 
applied to the fermion fields at $s=-1$ and $L_s$. For the 
sake of simplicity, we omit the Pauli-Villars
fields which can be introduced to cancel the bulk of the
heavy modes of the theory. The interested reader can 
consult Ref. \cite{shamir} for details.

As shown in Ref. \cite{shamir}, the above construction
in the absence of gauge fields has a low-energy effective
theory with two massless chiral
fermions near $s=0$ and $L_s-1$ in the limit of $L_s\rightarrow 
\infty$. Thus, a single-flavor massless Dirac field $q(x)$ 
can be constructed from the domain-wall fermion field 
$\psi(x, s)$ as 
\begin{equation}
     q(x) = P_L\psi(x, s=0) + P_R\psi(x, s=L_s-1) \ . 
\end{equation}
An explicit fermion mass, $m_f$, for the chiral modes 
can be introduced through the modified boundary conditions,
\begin{equation}
 P_L\psi_{x,L_s}= -m_fP_L\psi_{x,0}\ , \quad P_R\psi_{x,-1}
=-m_fP_R\psi_{x,L_s-1} 
 \ . \nonumber \\
\end{equation}
Substituting this into Eq. (\ref{action}), we obtain a generic
fermion mass term. 

To study the low-energy aspects of the 
domain-wall fermion action, we introduce the hermitian domain-wall 
Dirac operator, $H_{\rm DWF}(m_0) = R_{s}\gamma_5D_{\rm DWF}(m_0)$, 
where $R_s$ denotes the reflection in the fifth direction.
Because of the low-energy property mentioned
above, we expect that the low-lying spectrum of
$H_{\rm DWF}(m_0)$ resembles that of a massless Dirac particle and 
contributes dominantly to low-energy physical
observables. [Unlike the four-dimensional
Wilson-Dirac operator $D_{\rm W}(m_0)$ which has a simple 
hermitian-conjugation property, we find that the domain-wall 
Dirac operator $D_{\rm DWF}(m_0)$ has a peculiar spectrum 
with obscure physical significance.]

\begin{figure}[t]
\begin{center}
\vspace{-0.3in}
\epsfig{file=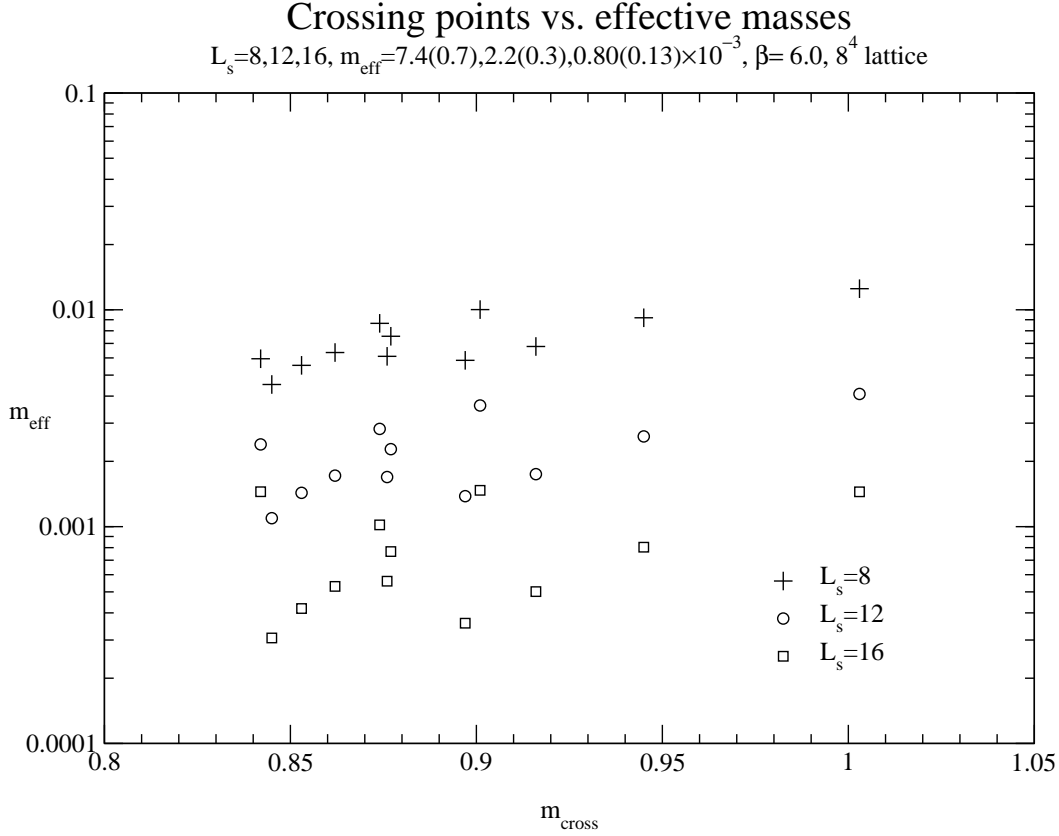,clip=,height=16cm,angle=-90}
\end{center}
\caption{Effective quark mass induced by domain walls 
for the Monte Carlo configurations at $\beta=6.0, 8^4$ lattice.
$L_s$ is the number of lattice sites in the fifth 
direction.}
\label{fig:60_1}
\end{figure}   

A more direct way of studying the low-energy physics 
of the domain-wall fermion is to integrate
out all heavy modes. This has been done by 
a number of authors \cite{neuberger,kikukawa,borici,edwards1}. 
The result has a very simple form
\begin{equation}
     S^{\rm eff} = \sum_{x,y} 
   \bar q(x) D^{\rm eff}_{L_s}(a_5,m_0) q(y) \ , 
\end{equation}
where we have made the lattice spacing in the
fifth direction $a_5$ explicit and 
\cite{edwards1}
\begin{eqnarray}
    D^{\rm eff}_{L_s}(a_5,m_0) &=& {1 + \gamma_5 \tanh 
    \left({{L_s\over 2}H(a_5,m_0)}\right)
     \over 1-\gamma_5 \tanh \left( {{L_s\over 2} H(a_5,m_0)} 
    \right)} \ , 
    \nonumber \\
    H(a_5,m_0) &=& {1\over a_5}\ln 
   \left({1+a_5P_RH_{\rm W}(m_0) \over 1-a_5P_LH_{\rm W}(m_0)} 
     \right)  \ . 
\label{relation}
\end{eqnarray}
$H_{\rm W}(m_0) = \gamma_5 D_{\rm W}(m_0)$ is the hermitian 
Wilson-Dirac operator. The four-dimensional effective Dirac operator 
$D^{\rm eff}_{L_s}(a_5,m_0)$ can be used to calculate the propagator
of the chiral fermion modes. If an observable involves
the fermion determinant, we must take into
account the contribution from the Pauli-Villars particles as well. 
The effective Dirac operator in the fermion determinant
is
\begin{equation}
     \overline{D}_{L_s}(a_5,m_0) 
    = {1\over 2}\left[1+\gamma_5 \tanh\left({{L_s\over 2}
     H(a_5,m_0)}\right)\right]\ . 
\end{equation}
In the limit of a continuous fifth dimension, i.e. 
$a_5\rightarrow 0$, $H(a_5,m_0)$ in the above expression becomes
\begin{equation}
      H(a_5=0,m_0) = H_{\rm W}(m_0) \ . 
\end{equation} 
And if we take the limit $L_s\rightarrow \infty$ as well, 
\begin{equation}
    \tanh\left({L_s\over 2}H\right) \rightarrow \epsilon(H) 
   = \theta(H)-\theta(-H) \ , 
\label{ep}
\end{equation}
the Dirac operator $\overline{D}_\infty(a_5=0,m_0)$ 
becomes the well-known 
Neuberger overlap operator \cite{neuberger}. On the other
hand, if we take $L_s\rightarrow \infty$ first, we have
\begin{equation}
    \overline{D}_\infty(a_5,m_0) 
   = {1\over 2}\left[1+ \gamma_5 \epsilon(H(a_5,m_0))\right] \ . 
\end{equation}
The difference from the Neuberger overlap operator 
\cite{neuberger} is 
that $H(a_5,m_0)$ is now a function of $H_{\rm W}(m_0)$ (see Eq. (\ref{relation})).  

\begin{figure}[t]
\begin{center}
\vspace{-0.3in}
\epsfig{file=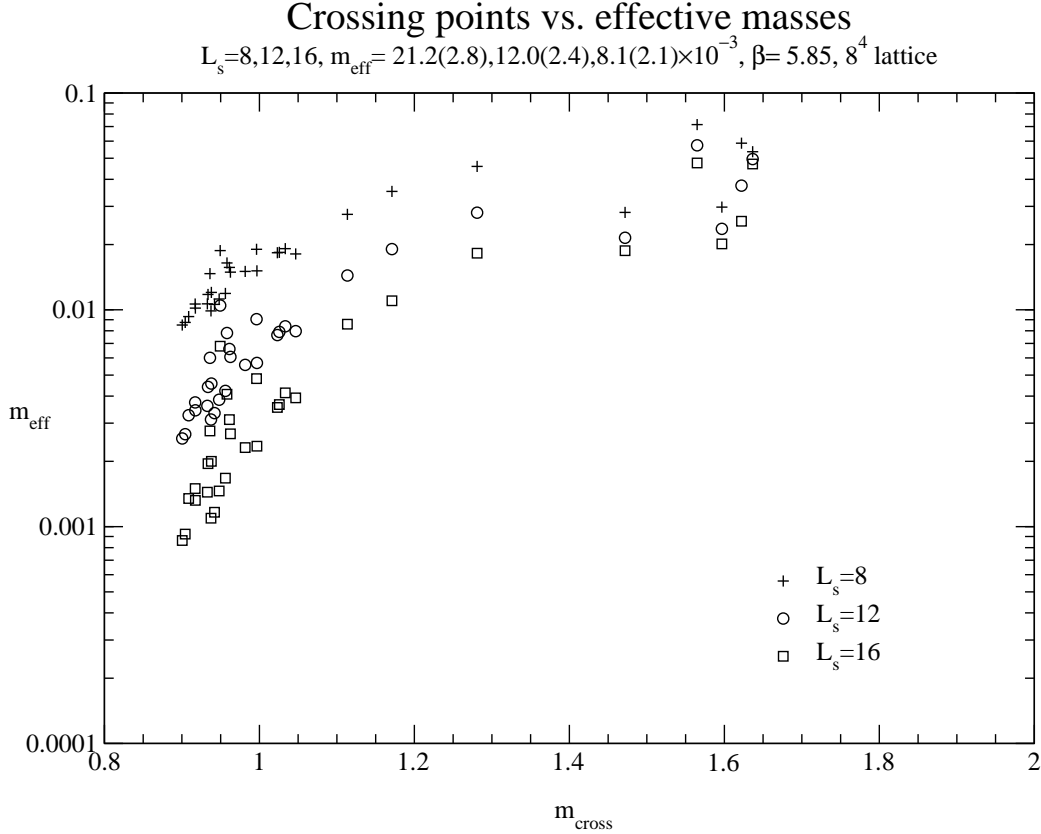,clip=,height=16cm,angle=-90}
\end{center}
\caption{Effective quark mass induced by domain walls 
for the Monte Carlo configurations at $\beta=5.85, 8^4$ lattice.
$L_s$ is the number of lattice sites in the fifth 
direction.}
\label{fig:585_2_1}
\end{figure}   

The above discussion leads to a simple and natural way  
of defining the topology of lattice gauge configurations
\cite{overlap}. 
The effective domain-wall Dirac operator at $L_s\rightarrow\infty$ 
($\overline{D}_\infty(a_5,m_0)$) has the following 
interesting property. If the hermitian operator
$H(a_5,m_0)$ has unequal numbers of positive 
and negative eigenvalues, its determinant vanishes; 
this means that $\overline{D}_\infty(a_5,m_0)$
has zero-energy eigenvalues. Inspired by the 
Atiyah-Singer theorem in the continuum spacetime, one can 
regard the appearance of the zero eigenvalues as 
the signal for a nontrivial topology of the lattice configuration. 
Since the number of positive and negative levels are the same 
for $m_0<0$, the difference between the number of positive and 
negative levels at a positive $m_0$ can be calculated 
by tracking the spectral flow of all eigenvalues 
of $H(a_5,m)$ between $m=0$ and $m=m_0$ and 
locating the level crossings 
\cite{overlap,specflow1,specflow2}. 
The topological index of a gauge configuration is given 
by the difference of the number 
of positive and negative energy levels of $H(a_5,m_0)$. 
The effective domain-wall 
Dirac operator $\overline{D}_{\infty}(a_5,m_0)$ has the same 
number of zero eigenvalues as the topological index. 
Rather than find the eigenvalues of $H(a_5,m)$, a computationally
simpler method can be used for $m<2$ and $a_5=1$ where the spectral
flow of the lowest eigenvalues of the hermitian Wilson-Dirac operator
$H_{\rm W}(m)$ are tracked between $m=0$ and $m=m_0<2$. The
topological index thus produced is the same as that of the effective
domain-wall Dirac operator $\overline{D}_{\infty}(a_5,m_0)$ \cite{edwards1}.

For finite $L_s$, however, the chiral modes of  
$\overline{D}_{L_s}(a_5,m_0)$ corresponding
to the gauge field topology do not have 
exact zero eigenvalue. The limit is approached 
exponentially as $L_s\rightarrow \infty$. Since the 
low-lying eigenvalues of the effective Dirac operator 
$\overline{D}_{L_s} (a_5,m_0)$ must reproduce those of 
$H_{\rm DWF}(m_0)$, the latter has the same 
number of exponentially-small eigenvalues as 
the topological index. We {\it define} the
finite-$L_s$ induced fermion mass $m_{\rm eff}$ as
an average of these quasi-zero eigenvalues. 

In a previous publication \cite{gjj}, we have calculated 
$m_{\rm eff}$ from 150 configurations on an $8^4$ lattice
at $\beta=6.0$ and $m_0=1.8$. Here again we show $m_{\rm eff}$ for each of
the zero modes in Fig. \ref{fig:60_1} where the horizontal
axis is the Wilson mass $m$ representing the
locations of the level crossings in the eigenvalue 
flow of $H_{\rm W}(m)$, and the vertical axis is $m_{\rm eff}$. 
For those configurations with multi-level crossings, we relate
$m_{\rm eff}$ and $m$ by assuming that the spectrum flow 
is repulsive or levels do not cross, and this simplifies 
the presentation. The results for three different 
$L_s$ are shown with three different symbols. 

How does the effective mass change as the coupling grows
stronger (smaller $\beta$)? To answer this question, 
we have generated 100 $SU(3)$ 
lattice gauge configurations on a lattice size $8^4$, 
50 each at $\beta =5.85$ and 5.7. We measure the 
eigenvalue flow of the hermitian Wilson-Dirac 
operator, and calculate the eigenvalues of 
$H_{\rm DWF}(m_0)$ corresponding to the nontrivial 
topology of the gauge configurations. For larger lattice
spacing, quantum fluctuations are stronger, and some of
these fluctuations can be misidentified as small size 
instantons. It turns out that they can induce strong
couplings between the left-and right-handed chiral modes 
and are detrimental to the existence of the low-energy 
effective theory. Indeed, for the same $L_s$, the effective 
masses are much larger at the smaller $\beta$'s 
than those at $\beta =6.0$. For example, with $L_s=16$, 
$m_{\rm eff}$ is 0.0008 at $\beta=6.0$, 0.008 at $\beta=5.85$, 
and 0.018 at $\beta = 5.7$.  

\begin{figure}[t]
\begin{center}
\vspace{-0.3in}
\epsfig{file=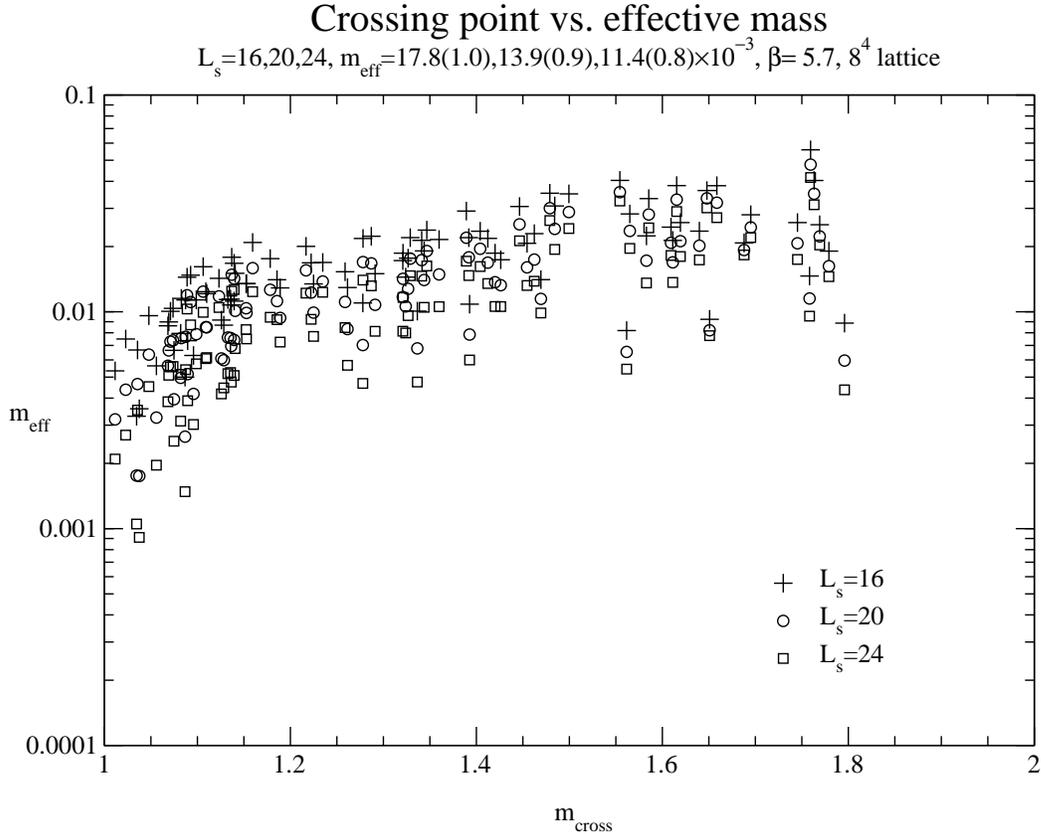,clip=,height=16cm,angle=-90}
\end{center}
\caption{Effective quark mass induced by domain walls 
for the Monte Carlo configurations at $\beta=5.7, 8^4$ lattice.
$L_s$ is the number of lattice sites in the fifth 
direction.}
\label{fig:57_1}
\end{figure}

Moreover, for the same lattice size, 
physical volume is larger at smaller $\beta$, and hence 
can house more instantons. 
As shown in Fig. \ref{fig:585_2_1}
for $\beta=5.85$ and Fig. \ref{fig:57_1} for 
$\beta = 5.70$, the total number of instantons 
in the 50 configurations is now 32 and 96, 
respectively. The level crossings happen at 
larger $m_0$ compared with Fig. 1. For $\beta=5.7$, 
there are several crossings very close to $m_0=1.8$. 
This is in contrast to what has been observed 
at $\beta=6.0$, where the crossings occur mostly
around $m=1.0$ \cite{gjj,specflow2}. 
The equal spacing between $m_{\rm eff}$ at different 
$L_s$ is a clear signal for the exponential 
decay. However, there is a significant 
variation in the rate among all the crossings, 
as evident in the figures. (Note: After the analysis 
for the paper is completed, the recent work of the CP-PACS \cite{cppacs}
 and RBC \cite{rbc} collaborations became available, which shows
the signal for varying rate of exponential decay and/or
nonvanishing effective mass in the $L_s \rightarrow \infty$ limit.
This behavior is only seen at a much larger $L_s$ than those studied here.
However, it is interesting to note that averaging eigenvalues with 
varying exponential rate can easily reproduce the large $L_s$ behavior of the effective
mass observed in the aforementioned reference.)

\begin{figure}[t]
\begin{center}
\vspace{-0.3in}
\epsfig{file=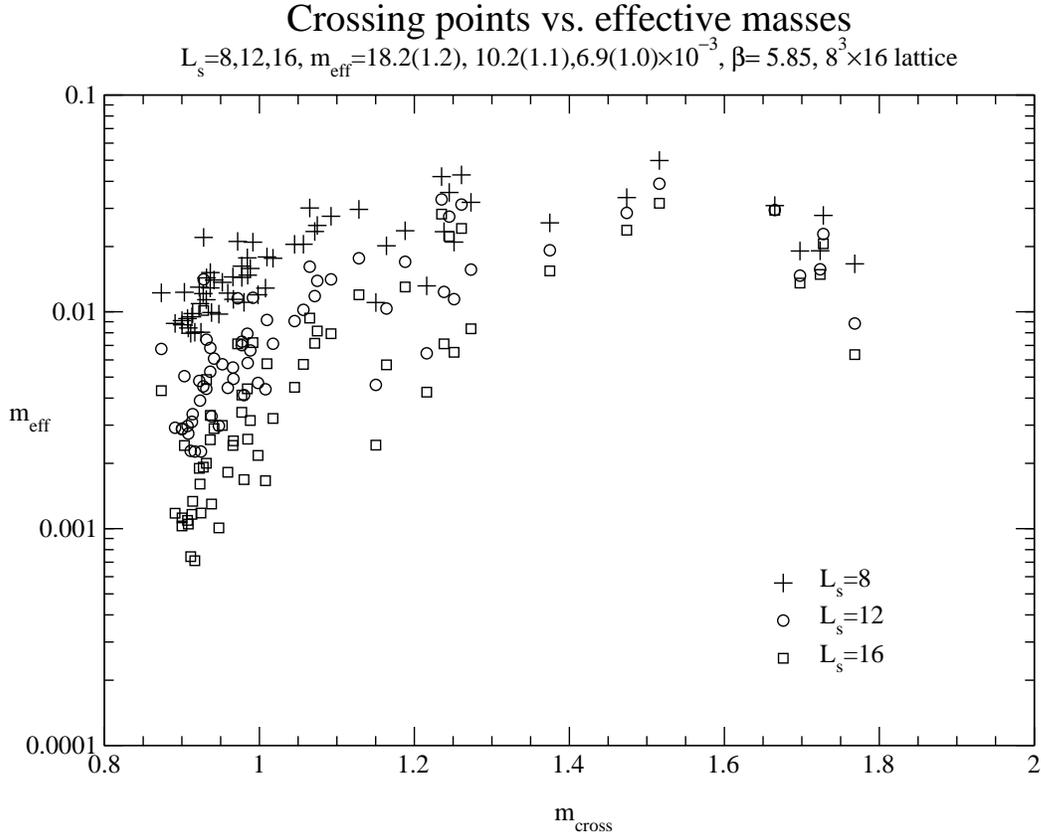,clip=,height=16cm,angle=-90}
\end{center}
\caption{Effective quark mass induced by domain walls 
for the Monte Carlo configurations at $\beta=5.85, 8^3 \times 16$ lattice.
$L_s$ is the number of lattice sites in the fifth 
direction.}
\label{fig:585_1}
\end{figure}   

To see the volume dependence  
at a fixed $m_0$ and $\beta$, we also measure the effective
mass on a set of 50 configurations on an $8^3\times 16$ lattice 
at $\beta=5.85$. The total number of instantons 
is now 64, doubling that on the $8^4$
lattice. We notice there are several crossings
very close to $m_0=1.8$. The average effective
mass turns out to be essentially the same as that 
on the smaller lattice. A similar conclusion can be drawn by 
comparing $m_{\rm eff}$ at $\beta=6.0$
and $V=8^4$ with that at the same $\beta$ and $V=16^3\times 24$
in Ref. \cite{aoki}, although we note 
that the way of determining the effective
mass there is quite different. 
This weak dependence on the size of the volume
may allow us to extrapolate our quantitative results to 
large lattices necessary for realistic simulations.

\begin{figure}[t]
\begin{center}
\vspace{-0.3in}
\epsfig{file=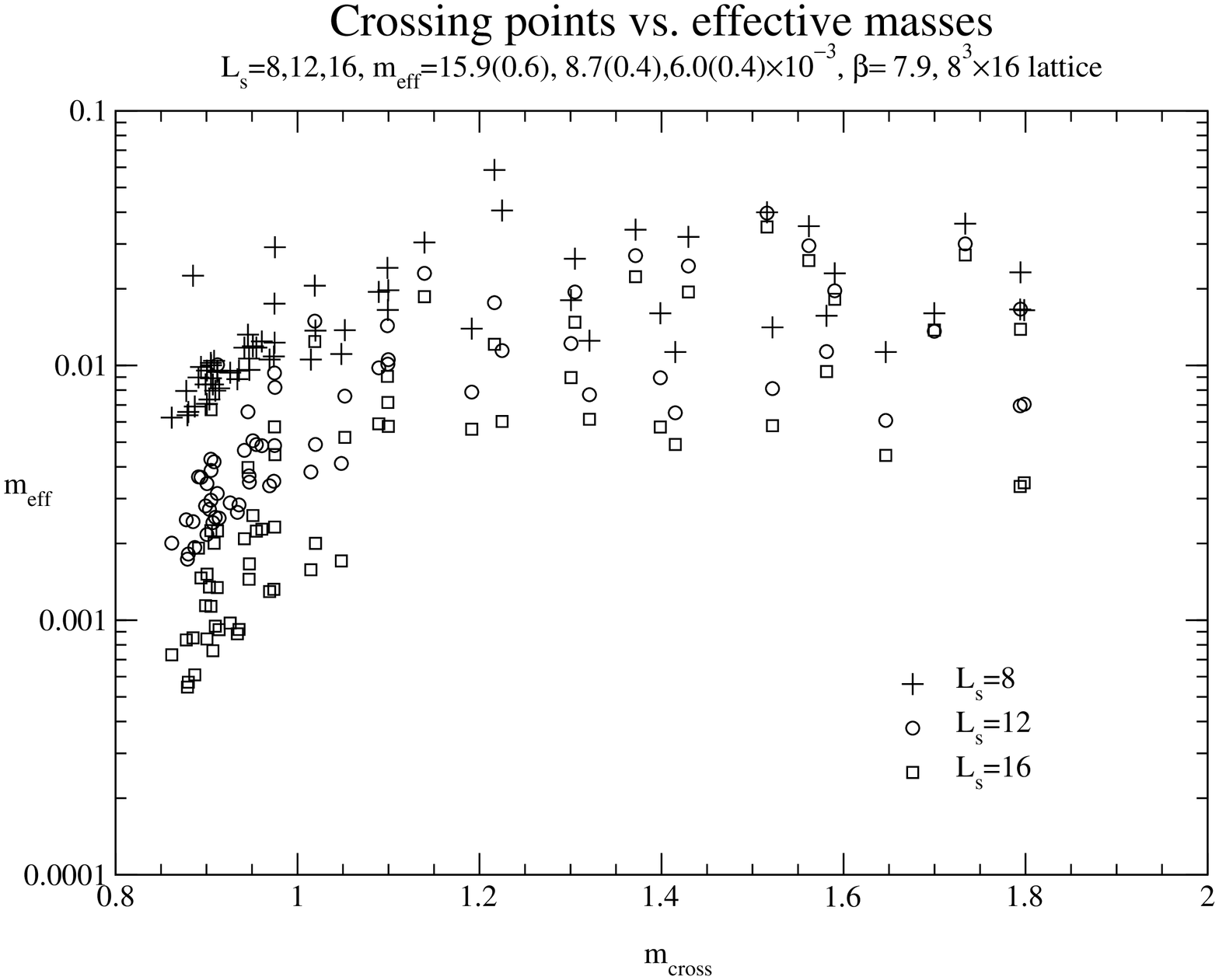,clip=,height=16cm,angle=-90}
\end{center}
\caption{Effective quark mass induced by domain walls 
for the 50 Monte Carlo configurations generated by the 1-loop, tadpole
improved gauge action \cite{LW} on a $\beta=7.9, 8^3 \times 16$ lattice.
$L_s$ is the number of lattice sites in the fifth 
direction.}
\label{fig:LW}
\end{figure}   

We also measure the effective mass 
on a set of 200 configurations generated from 
the 1-loop L\"uscher-Weisz gauge action \cite{LW} with
the tadpole improvement. (Crossings from only 50 lattices are shown in Fig.
\ref{fig:LW}.) Similar studies using various improved gauge actions,
including  an RG improved action
\cite{iwasaki} are reported in \cite{cppacs,vranas}. 
The gauge coupling ($\beta= 7.9$) corresponds to a 
spacing of $\sim$ 0.16 fm, similar to 
$\beta=5.7$ of the Wilson action. The spectral flow
and the domain-wall eigenvalues are studied with the same 
Wilson-Dirac operator. 
As shown in Fig. \ref{fig:LW}, 
the number of instantons as well as the distribution 
of the crossings differs significantly from the 
Wilson action at similar lattice spacing.
Because of the decrease of small-scale 
quantum fluctuations, the probability of the 
crossing at $m_0 >1.2$ is heavily suppressed, 
and the density of small eigenvalues of $H_{\rm W}(m_0)$ 
is also much smaller ($\sim 2.7 \times 10^{-4}$, compared to 
$\sim 1.8 \times 10^{-3}$ for $\beta=5.7$ Wilson action). 
Therefore, the effective mass 
decreases faster as a function of $L_s$. 
The average effective mass at $L_s=16$ is $\sim 6 \times
10^{-3}$, compared to $18 \times 10^{-3}$ from
the Wilson action at $\beta=5.7$.

\begin{figure}[t]
\begin{center}
\vspace{-0.3in}
\epsfig{file=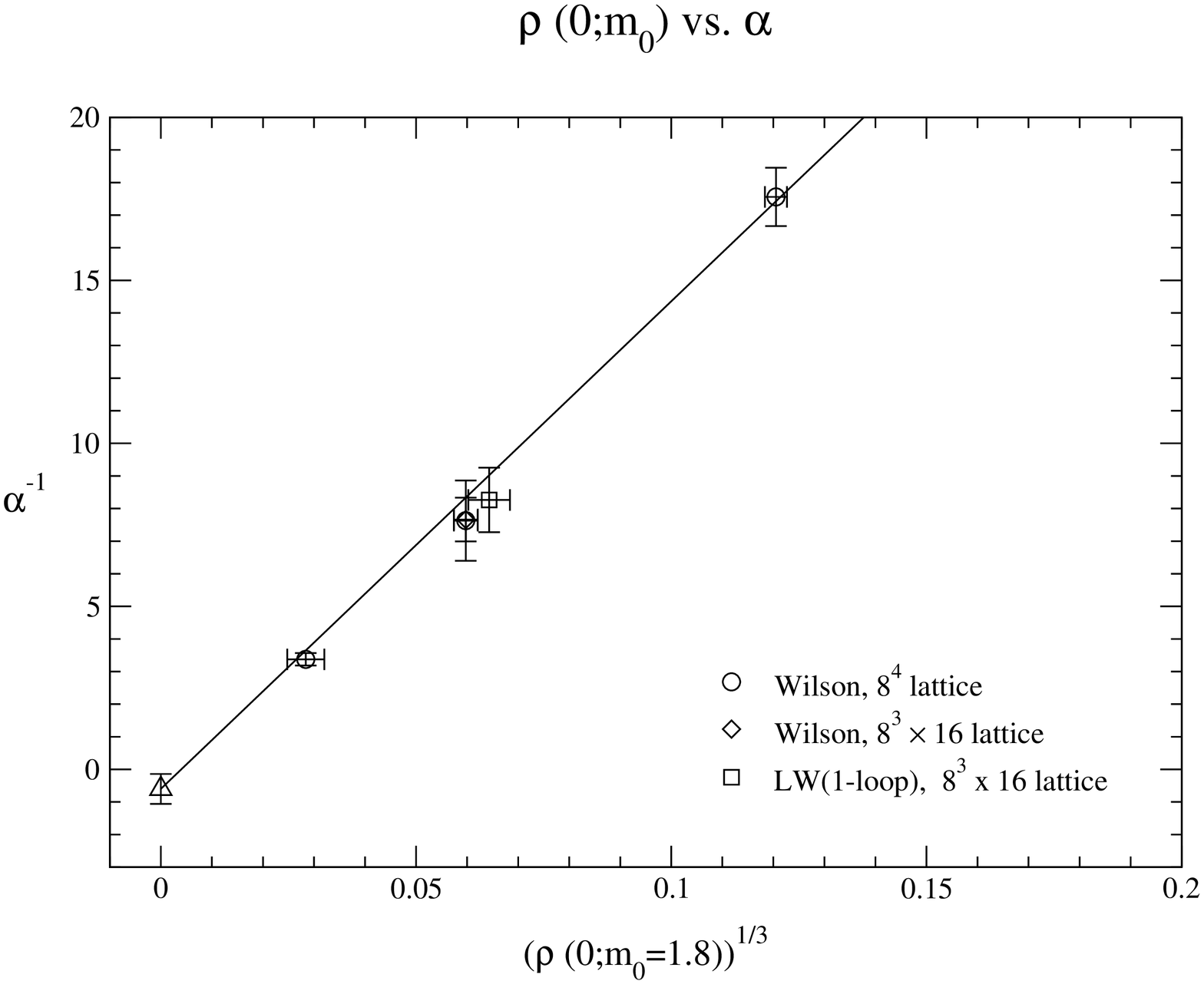,clip=,height=16cm,angle=-90}
\end{center}
\caption{Average coefficient of the exponential decreases
as a function of $\rho(0;m_0)$. The extrapolation of the fit to the
continuum limit for the Wilson gauge backgrounds is shown on the left.
$\rho(0;m_0)$ for the Wilson and improved gauge action are from 
Ref. \cite{specflow2} for $8^3\times 16$ lattices.} 
\label{fig:rho_0}
\end{figure}

The origin of the induced chiral fermion mass is 
the finiteness of $L_s$. More specifically, the hyperbolic-tangent 
function at finite $L_s$ is used to approximate  
the $\epsilon$ function (Eq. (\ref{ep})). 
This approximation is good for large eigenvalues
of $H(a_5,m_0)$, but poor for small ones. For smaller
$\beta$, $H(a_5,m_0)$ has more small eigenvalues and hence
the hyperbolic-tangent is a worse approximation of the 
$\epsilon$-function. 
This point is also reflected in 
the dependence of the exponential decay rate on the density
of the zero eigenvalues of $H_{\rm W}(m_0)$ ($\rho(0;m_0)$). 
We note that since the gauge fields are replicated along the fifth
dimensional slices, the relevant length scale in the fifth dimension
is the inverse of the rate of exponential decay ($\alpha$) and by simple 
engineering dimensions is given qualitatively by the density of 
zero eigenvalues of $\rho(0;m_0)^{1/3}$. 
In Fig. \ref{fig:rho_0}, we have plotted $1/\alpha$ 
as a function of $\rho(0;m_0)$.
The rate $\alpha$ for each coupling is calculated by fitting 
$m_{\rm eff} = m_0\exp[-\alpha L_s]$ at different $L_s$. 
The statistical errors are estimated by doing
correlated fits to single-eliminated jackknife blocks.
The inverse decay rates from all the configurations studied
show an approximate linear scaling as $\rho(0;m_0)^{1/3}$.
This suggests that the density of small eigenvalues is indeed 
the dominating factor for the exponential decay
rate of domain-wall fermions.




\begin{figure}[t]
\begin{center}
\vspace{-0.3in}
\epsfig{file=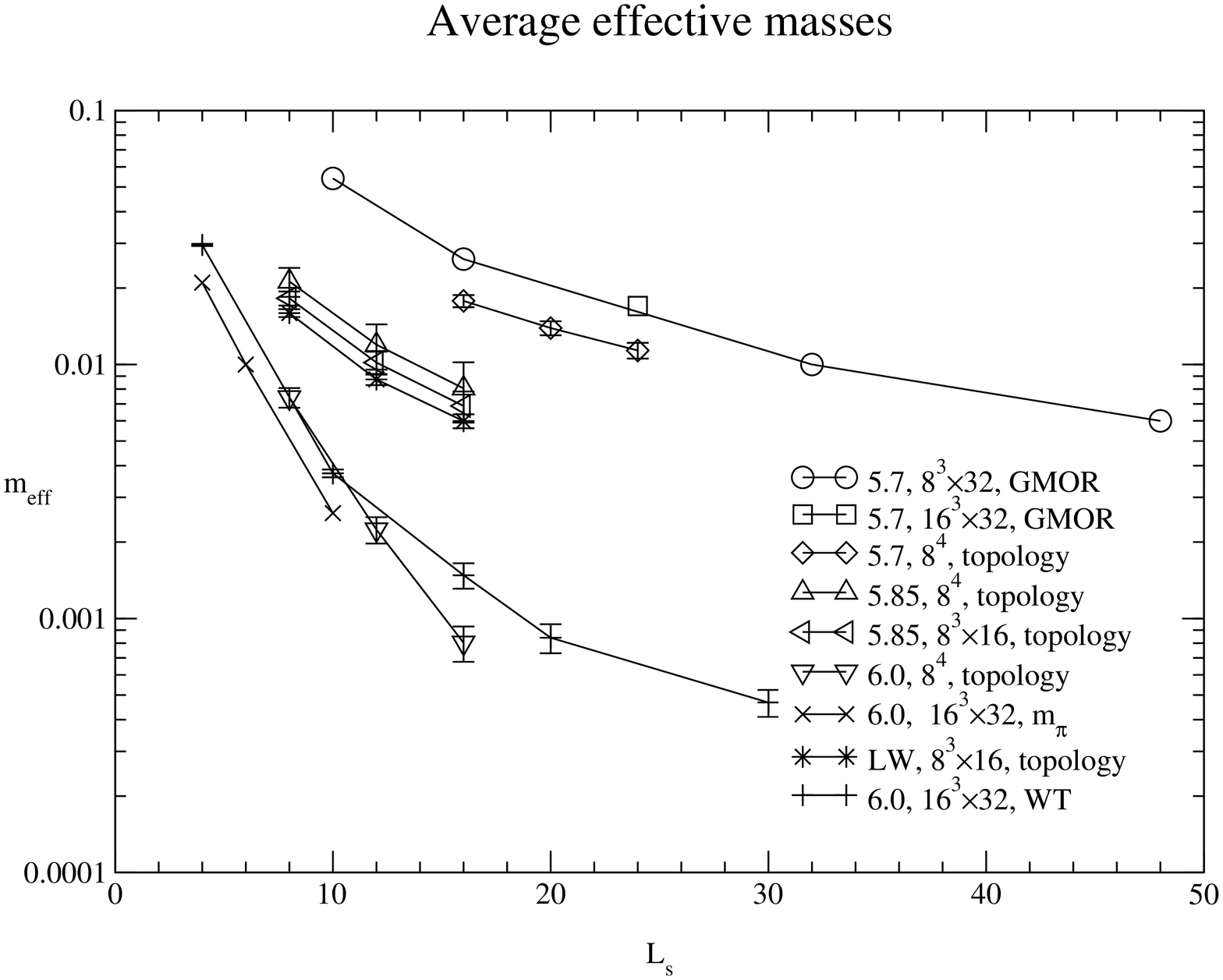,clip=,height=16cm,angle=-90}
\end{center}
\caption{Average effective masses from various observables as a function of
$L_s$. Effective masses from $\beta=5.7$, GMOR relation are from Ref. \cite{fleming}.
Data for $\beta=6.0$, $m_{\pi}$ and the axial WT identity are from Ref. \cite{aoki} and
\cite{cppacs}, respectively.
LW denotes the 1-loop, tadpole improved gauge action from
Ref. \cite{LW}.}
\label{fig:meff}
\end{figure} 

Effective masses thus obtained for the different gauge 
coupling and $L_s$ are plotted in Fig. \ref{fig:meff}. 
The data from Ref. \cite{aoki, fleming} are also included 
for comparison. 
For a given gauge coupling, different volumes
and methods of measurements have 
little effect on the size of the effective mass as well as
the rate of exponential decay. However, 
the change of gauge action affects the effective 
mass significantly. 
 It is quite clear from the figure \ref{fig:meff}
that for a practical simulation of the domain-wall
fermion, one either chooses a large $\beta$ with
the conventional Wilson action or an improved
action keeping lattice spacing large.

To summarize, we have studied the residual chiral 
symmetry breaking present in domain-wall fermion by 
measuring the eigenvalues of the hermitian domain-wall 
Dirac operator corresponding to the topology of
the lattice gauge configurations. Individual eigenvalues 
for the topological zero modes show clear exponential
behavior in $L_s$. We regard these eigenvalues as
the induced mass for the surface chiral modes at finite
$L_s$ separation. 

For $L_s$ and $\beta$, we see little variation
of $m_{\rm eff}$ as a function of the volume. This is 
in some sense expected because the coupling of the chiral
modes between the opposite walls has little to do 
with the size of the four-dimensional slice. On the
other hand, a strong dependence on $\beta$ is observed.
In particular, the effective mass is much larger 
at $\beta=5.85$ or 5.7 than that at $\beta=6.0$. 
For the improved gauge action, the 
spurious fluctuations are reduced significantly
 and the $L_s$ needed to obtain a good chiral symmetry
is reduced. Since the additional computation needed for 
the improved action is negligible,
using improved gluon actions may enable us to simulate
domain-wall fermions with larger lattice spacing.

\acknowledgments
We thank N. Christ and J. Negele for useful
discussions related to the subject of this paper. 
The numerical calculation reported here
were performed on the Calico Alpha Linux Cluster and the QCDSP
at the Jefferson Laboratory, Virginia.
C.J., X.J. and V.G. are supported in part by funds provided by the
U.S.  Department of Energy (D.O.E.) under cooperative agreement
DOE-FG02-93ER-40762. R.G.E. was supported by DOE contract DE-AC05-84ER40150 
under which the Southeastern Universities Research Association (SURA) 
operates the Thomas Jefferson National Accelerator Facility (TJNAF).

\end{document}